\documentclass[prl,twocolumn,preprintnumbers,floatfix,nofootinbib,letterpaper]{revtex4}

\input{epsf}
\usepackage{epsf}
\usepackage{amssymb,graphics,graphicx}
\usepackage{footnote}
\usepackage{natbib}

\begin{document}

\title
{\Large \bf Pseudo-rip: Cosmological models intermediate between
the cosmological constant and the little rip}
\author{Paul H. Frampton$^{1}$, Kevin J. Ludwick$^{1}$, and
Robert J. Scherrer$^{2}$}
\affiliation{$^1$Department of Physics \& Astronomy, University of North
Carolina, Chapel Hill, NC~~27599}
\affiliation{$^2$Department of Physics \& Astronomy, Vanderbilt University,
Nashville, TN~~37235}
\date{\today}

\begin{abstract}

If we assume that the cosmic energy density will remain constant or strictly increase 
in the future, then the possible fates
for the universe can be divided into four
categories based on the time
asymptotics of the Hubble parameter $H(t)$:  the cosmological
constant, for which $H(t) =  constant$, the big rip, for which $H(t) \rightarrow
\infty$ at finite time, the little rip, for which $H(t) \rightarrow \infty$
as time goes to infinity, and the pseudo-rip, for
which $H(t) \rightarrow constant$ as time goes to infinity.
Here we examine the last of these possibilities in more detail.
We provide models that
exemplify the pseudo-rip, which is an intermediate
case between the cosmological constant and the little rip.
Structure disintegration in the pseudo-rip depends on the model parameters.
We show that pseudo-rip models for which the density and Hubble parameter
increase monotonically can produce an inertial force which
does not increase monotonically, but instead peaks at a particular
future time and then decreases.

\end{abstract}

\pacs{}\maketitle

{\it Introduction}---
Time is an elusive concept of great
interest to all physicists since it underlies
all dynamical systems. Although we can measure time
on everyday scales with exquisite precision,
the deeper meaning of time
may be addressed only by study
of the origin and fate of the universe.

Current observations \cite{perlmutter,kirshner,union08,hicken} 
strongly suggest that the universe is dominated by
a negative-pressure component, dubbed dark energy.  This component
can be characterized by an equation of state parameter $w$, which is
simply the ratio of the pressure to the density:  $w = p/\rho$; for example,
a cosmological constant corresponds to $w = -1$.  While
it is often assumed that $w \ge -1$ in accordance
with the weak energy condition, it has long been known
\cite{Caldwell} that the observations are consistent with $w < -1$,
which corresponds to a dark energy density that increases with
time $t$ and scale factor $a$.  If the density increases monotonically in
the future, then the universe can undergo a future singularity, called
the ``big rip," for which
$\rho \rightarrow \infty$ and $a \rightarrow \infty$
at a finite time.  Shortly before this singularity is reached, bound structures
are disintegrated by the expansion \cite{CKW}.

Note, however, that a dark energy component with a monotonically increasing
density that is unbounded from above does not lead 
inevitably to a future singularity, although it
does ultimately lead to the dissolution of all bound structures.  Such models, dubbed
``little rip" models, were first examined in detail by Frampton et al.
\cite{FLS1},
who derived the boundary
between big rip and little rip models
in terms of
$\rho(a)$.  Properties
of little rip models were
further investigated in Ref. \cite{FLNOS}.

Here we investigate a different set of models, in which the density of the dark energy
increases monotonically with scale factor but is bounded from above by
some limiting density, $\rho_\infty$.  Such models can still lead
to a dissolution of bound structures for a sufficiently strong inertial force (which we will define), 
so we dub these models ``pseudo-rip" models.  This allows us, in the context of monotonically
increasing $\rho$, to distinguish three distinct cosmic futures:  the big rip, little rip,
and pseudo-rip.  (Note that models for which $\rho(a)$
is not monotonic are physically less plausible, and it is more difficult to make
any sort of general statement about such models).
While
the dissolution of bound structures is inevitable in the big rip and little rip scenarios,
it may or may not occur in a pseudo-rip, depending on the model parameters.

In the next section, we provide the definition of the pseudo-rip and examine
the conditions necessary to dissolve bound structures.  We then examine two specific 
functional forms of the dark energy density and discuss scalar field realizations of pseudo-rip 
models.  We assume a flat FLRW metric and $c=1$ throughout.

{\it Definition of Pseudo-Rip}---Before defining the pseudo-rip, we suggest a compact
way to classify all rips. 
We find it useful to classify ripping behaviour
by means of the Hubble parameter $H(t)$. By ripping
behavior, we mean any future evolution which can
lead to disintegration of structure in the form of bound
systems by virtue of the strong inertial force due to dark energy.

Given the Hubble parameter $H(t)$ for $t \geq t_0$, where 
$t_0$ is the present time, the density $\rho(t)$ and pressure $p(t)$ are:
\begin{equation}
\rho(t) = \left( \frac{3}{8 \pi G} \right) H(t)^2,
\label{Friedmann}
\end{equation}
\begin{equation} 
p(t) = - \left( \frac{1}{8 \pi G} \right) \left[2 \dot{H}(t) + 3 H(t)^2 \right].
\label{pressure}
\end{equation}

The big rip is defined by
\begin{equation}
H(t) \longrightarrow + \infty,  ~~~~~ t \longrightarrow t_{rip} < \infty. 
\label{Bigrip}
\end{equation}
In a big rip, all bound structures dissociate in a finite time in the future, and 
space-time ``rips apart'' at a finite time in the future, i.e., the scale factor 
of the FLRW metric goes to infinity at $t = t_{rip}$ \cite{CKW}. The little rip is defined by
\begin{equation}
H(t) \longrightarrow +\infty, ~~~~~  t \rightarrow +\infty.
\label{Littlerip}
\end{equation}
The little rip dissociates all bound structures, but the strength of the dark energy 
is not enough to rip apart space-time as there is no finite-time singularity.

An excellent fit to all cosmological data is
provided by the $\Lambda$CDM model which, in present parlance,
is a ``no-rip'' model defined by
\begin{equation}
H(t) = H(t_0). 
\label{LammdaCDM}
\end{equation}

There remains just one additional possibility
for monotonically increasing $H(t)$, namely
\begin{equation}
H(t) \longrightarrow H_\infty < \infty, ~~~~~ t \rightarrow +\infty,
\label{pseudorip}
\end{equation}
where $H_\infty$ is a constant.  Equation (\ref{pseudorip}) defines
the pseudo-rip, the subject of the
present article.  A pseudo-rip dissociates bound structures that are held together by 
a binding force at or below a particular threshold that depends on the
inertial force in the model.
Eqs.(\ref{Bigrip})-(\ref{pseudorip}) 
clearly exhaust all possibilities for a monotonically
increasing $H(t)$, i.e., a monotonically increasing
dark energy density
$\rho(a)$.

The equations for monotonically increasing $H(t)$ are the same for $\rho(t)$  {\it mutatis mutandis}.
Pressure $p(t) \longrightarrow -\rho(t)$ as $t \longrightarrow \infty$,
provided $ \dot{H}(t) \longrightarrow 0$ on the R.H.S. of
Eq.(\ref{pressure}), which is the case for a pseudo-rip model.

Our division of the future evolution into the categories of
big rip, little rip, and pseudo-rip represents a different set of models than
those
examined in Ref. \cite{NOT}, which provided a classification scheme
for future singularities.
Our scheme represents a classification of all models with monotonically
increasing dark energy density, for which
the scale factor $a$ goes
smoothly to infinity, at either finite or infinite time, and for which
there are no singularities in the derivatives of $H$ unless $H$ itself
is singular.  In our scheme, the big rip encompasses the type I singularity
of Ref. \cite{NOT}, while the type II, III, and IV singularities lie outside
the types of models considered in our classification scheme.
The little rip and pseudo-rip models are by definition non-singular, so they 
fall outside 
of the purview of Ref. \cite{NOT}.

The inertial force $F_{inert}$ on a mass $m$ as seen by a gravitational source 
separated by a comoving distance $l$ is given by \cite{FLNOS}
\begin{eqnarray}
F_{inert} &=& m l ( \dot{H}(t) + H(t)^2) \nonumber\\
&=& -ml \frac{4 \pi G}{3} (\rho(a) + 3p(a)) \nonumber\\
&=& ml \frac{4 \pi G}{3} (2 \rho(a) + \rho '(a) a). \label{Finert}
\end{eqnarray}
For simplicity, 
we set the scale factor at the present time  $a_0=a(t_0)=1$.  A bound structure 
dissociates when the inertial 
force, dominated by dark energy, grows in the future to equal the force holding 
together the bound structure in question.  For a pseudo-rip, $F_{inert}$ is
asymptotically finite.

However, if the bound structure is massive 
enough to significantly affect the local space-time metric, it is not accurate to 
express $F_{inert}$ in terms of the FLRW metric.  A more accurate method and local metric 
is employed in \cite{rip2}, and we use their method to calculate the disintegration 
times for the Milky Way and the Earth-Sun system.

We analyze two psuedo-rip parameterizations for dark energy density, models 1 and 2,
each as a function of the scale factor $a(t)$ with other parameters.

{\it Model One}---Model 1 is defined by
\begin{equation}
\rho_{1}(a,B,f,s)= \rho_{0} \frac{\ln[\frac{1}{f+\frac{1}{a}} + \frac{1}{B}]^s}{\ln[\frac{1}{f+1} + \frac{1}{B}]^s},
\label{model1}
\end{equation}
where $\rho_0$  is the present value of the dark energy density.
Note that $\rho_1$ is normalized to be $\rho_{0}$ at $a=1$, which
we define to represent the present.  
Then $\rho_{1}(a, B, f, s)$ is a function of the scale parameter $a(t)$ and of
three other parameters
$B$, $f$, and $s$. It is most convenient to fix $f$ and $s$, which mostly control the strength of the rip,
and to keep $B$ as a free paramater for fitting the supernova data.

We fix $f$ and $s$ 
to specify how powerful the psuedo-rip 
should be.  The remaining free parameter is chosen to make a best fit to the latest 
supernova data from the Supernova Cosmology Project \cite{SCP} with a reduced $\chi^2$ of $ \simeq 0.98$.

As examples of bound states we consider the Milky Way (MW), the Earth-Sun system (ES),
the hydrogen atom (H atom), and the proton. The first two, MW and ES, are gravitationally
bound, while for the H atom and proton, we must carefully consider the electromagnetic and strong color forces
respectively. In all cases the dark energy density increases monotonically from $\rho_0$ at the present time
to an asymptotic value. Depending on the parameters $B$, $f$, and $s$, the inertial force
can successively disintegrate the MW, the ES, the H atom and the proton.

In different cases, some of these bounds systems will be 
disintegrated and not others. If none of 
them are disintegrated, we shall refer to such a 
pseudo-rip as a ``failed rip.''

\begin{figure}[t]
\begin{center}
\includegraphics[scale=0.60]{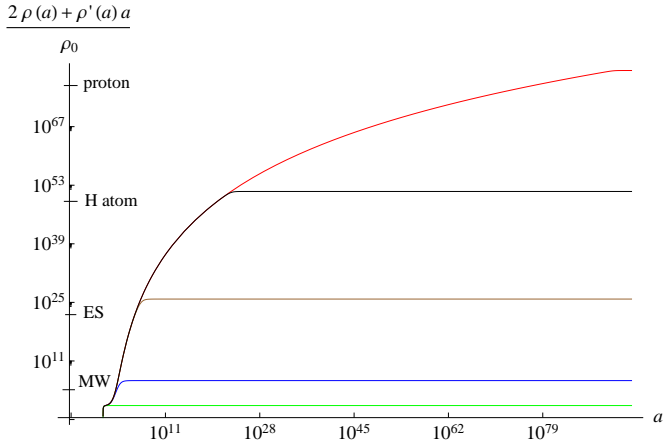}
\caption{Plotted from the innermost to the outermost 
curve is the scaled $F_{inert}$ for $\rho_1(a,0.29,10,48)$,
$\rho_1(a,0.0108,0.003,48)$, $\rho_1(a,0.01078,2 \times 10^{-7},48)$, 
$\rho_1(a,0.01078,10^{-23},48)$,
and $\rho_1(a,0.0108,10^{-92},48)$ respectively, for 
Model 1, given by Eq. (\ref{model1}).  
Each curve was fitted to supernova data with
the extra constraint that $B \geq 0$.  
The values necessary for structural disintegration are indicated.   
From the innermost curve
to the outermost:  failed rip; $t_{MW} - t_0 = 9.2 \times 10^4$ Gyr; 
$t_{ES}-t_0=1.6 \times 10^9$ Gyr;
$t_{H atom}-t_0=1.3 \times 10^{33}$ Gyr; $t_{proton}-t_0=3.8 \times 10^{117}$ Gyr.
(The color plots are in the online version of the paper.)}
\label{model1dis}
\end{center}
\end{figure}

In Fig. (\ref{model1dis}) we show five examples of the scaled 
$F_{inert}$ for $\rho_1(a)$, which 
include the matter and radiation contributions.  
Because of these contributions, the curves go 
to negative infinity as $a$ goes to $0$, and the 
$x$-intercepts are values of $a<1$ at which 
dark energy domination begins.  
Going from bottom to top,
the curves represent respectively a failed rip; 
a pseudo-rip which disintegrates only the MW;
a pseudo-rip which breaks apart the MW and ES; 
a pseudo-rip which destroys the MW, ES, and H atom; 
and the highest curve is for a pseudo-rip which succeeds 
in ripping apart all four of the 
MW, ES, H atom, and proton. Note that, unlike a little rip, the asymptotic value
of $\rho_{1}(a)$ as $a \longrightarrow \infty$ is finite.

For a given pseudo-rip model with monotonically increasing dark 
energy density $\rho_{DE}(a)$, 
the structures with bigger binding forces disintegrate 
after those with smaller binding forces.  But a 
particular $\rho_1$ can be constructed such that it leads 
to, for example, the disintegration of the proton 
before another $\rho_1$ disintegrates the Milky Way.

Notice that the more violent the pseudo-rip is required to be, 
the more extremely small the $f$ 
parameter is.  One may counteract this fine tuning of $f$ by, 
for example, introducing new 
factors into the model that help $\rho_1$ grow faster while 
still leaving it asymptotically finite.  
One such factor could be 
$(\frac{a+q}{a+2q} \frac{1+2q}{1+q})^w$.  Such a factor is $1$ when $a=1$, 
so $\rho_1$ is still $\rho_0$ at the present time.  
The new parameters $q$ and $w$ can 
avoid the fine tuning.

{\it Model 2}---Model 2 is defined by
\begin{equation}
\rho_{2}(a,A,n,m)= \rho_{0} \frac{A}{2} (\tan^{-1}(a-n) - \tan^{-1}(1-n)+1)^m.
\label{model2}
\end{equation}

\vspace{4 mm}
Like $\rho_1$, $\rho_2$ is normalized to be $\rho_{0}$ at $a=1$.
Then $\rho_{2}(a, A, n, m)$ is a function of the scale factor $a(t)$ 
and of three other parameters
$A$, $n$, and $m$. It is most convenient to fix $n$ and $m$, which 
mostly control the strength of the rip,
and to keep $A$ as a free paramater for fitting the supernova data.  
So we fix $n$ and $m$ 
to specify how powerful the psuedo-rip 
should be, and the remaining free parameter is chosen to make a best 
fit to the latest
supernova data from the Supernova Cosmology Project \cite{SCP} 
with a reduced $\chi^2$ of $ \simeq 0.98$.

\begin{figure}[h]
\begin{center}
\includegraphics[scale=0.60]{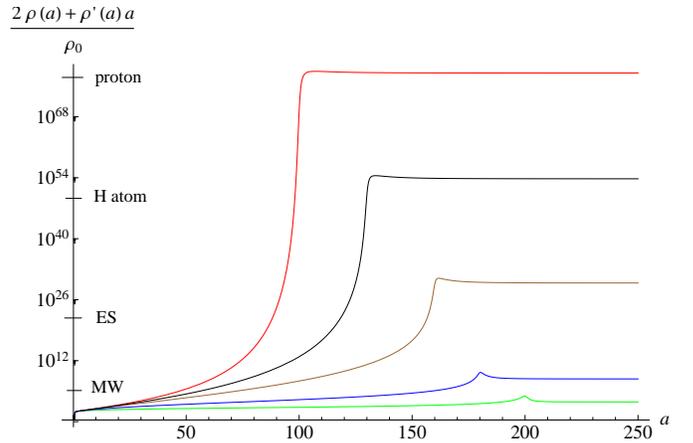}
\caption{Plotted from the innermost to the outermost curve 
is the scaled $F_{inert}$ for 
$\rho_2(a,17481.3,200,0.5)$, $\rho_2(a,3601.31,180,2)$, 
$\rho_2(a,571.1,160,10)$, $\rho_2(a,171.045,130,22)$,
and $\rho_2(a,55.45,100,40)$ respectively, for Model 2, 
given by Eq. (\ref{model2}).  Each curve was fitted to supernova data.  
The values necessary for structural disintegration are indicated.  
As explained in the text, 
disintegration times from different models should not be 
directly compared.  From the innermost curve
to the outermost:  failed rip; $t_{MW} - t_0 = 45$ Gyr; $t_{ES}-t_0=40$ Gyr;
$t_{H atom}-t_0=39$ Gyr; $t_{proton}-t_0=38$ Gyr.
(The color plots are in the online version of the paper.)}
\label{model2dis}
\end{center}
\end{figure}

In Fig. (\ref{model2dis}), we plot five examples of the 
scaled $F_{inert}$ for $\rho_2(a)$, 
which include the contributions from matter and radiation.  Just as in Fig. (\ref{model1dis}), 
the contributions cause the curves to approach negative infinity as $a$ goes to $0$, and the 
$x$-intercepts are values of $a<1$ at which dark energy domination begins.  
The bottom curve exhibits a failed rip, while the others show pseudo-rips of various 
strengths.  As was mentioned for Model 1, we see from 
Fig. (\ref{model2dis}) that the disintegration time for 
the proton for a particular $\rho_2$ can be sooner than 
the disintegration time for the Milky Way for another 
$\rho_2$.

Note in Fig. (\ref{model2dis}) that each $F_{inert}$ has a 
local maximum.  Because of this bump, $\rho_2(a \rightarrow 
\infty)$ is less than the maximum value of $\rho_2$.  So it 
is possible, for particular parameterizations, for 
$F_{inert}$ to reach the level of the binding force of a bound 
structure and then decrease, allowing 
the structure to possibly come back together.  In principle, 
a pseudo-rip model can have $F_{inert}$
with an arbitrary number of local maxima that give structures 
the chance to dissociate and reform multiple times.  
All this can be achieved using a functional form for 
dark energy density that is monotonically increasing.  
However, all the examples 
shown in Fig. (\ref{model2dis}) have their asymptotic 
values higher than the values necessary to 
rip apart the structures mentioned in the plot.

{\it Scalar Field Realizations}---One possible realization of 
pseudo-rip models is a minimally-coupled
phantom model, i.e., one that involves a scalar field with a negative kinetic term.
The equation of motion for such a field is
\begin{equation}
\label{phiev}
\ddot{\phi} + 3 H \dot{\phi} - V^\prime(\phi) = 0\, ,
\end{equation}
where the dot is a time derivative, and the prime denotes the derivative
with respect to $\phi$.  A field obeying this equation of motion rolls ``uphill"
in the potential.

It is clear that a sufficient condition for a pseudo-rip is that
$V(\phi) \rightarrow V_0$ (where $V_0$ is a constant) as $\phi \rightarrow
\infty$, and in Ref. \cite{FLNOS} it was shown that this is also
a necessary condition for a monotonic potential.
In this case, we simply have $\rho_\infty = V_0$.  If the potential
is not monotonic,
the density of the scalar field can also approach a constant
asymptotically if the field gets trapped in a local maximum with
$V = V_0$.  Phantom fields with bounded potentials have been discussed
previously in Ref.~\cite{Elizalde:2004mq}.

Note, however, that our discussion of pseudo-rips is much more general
than the specific example provided by phantom field models.  Phantom fields represent
only a single possible realization of this much more general class of models
for the asymptotic expansion of the universe.

{\it Discussion}---We have described merely two illustrative models 
of the pseudo-rip. Obviously,
there is an infinite number of possibilities.

A failed rip will disintegrate nothing because the inertial force will not
reach a high enough magnitude. Model 1 exemplifies a pseudo-rip model 
which can variously disintegrate between one and all four of the chosen systems
while $\rho_1$ asymptotes to a finite density, which implies a finite inertial force, 
as the scale factor $a(t)$
approaches infinity. As can be seen from Fig. (\ref{model1dis}), these quantities plateau
to a constant value after the last disintegration has taken place.

We included Model 2 as a particularly interesting example in which the 
inertial force rises relatively abruptly before it plateaus, 
as shown in Fig. (\ref{model2dis}). 
The various disintegrations
take place soon after each other in a cosmological sense.

For a function that asymptotically approaches a constant, such as $\rho_1(a)$ or 
$\rho_2(a)$, a point of inflection allows the function to have an 
arbitrarily large slope for a portion of the 
domain and still increase monotonically.  A point of inflection 
is present in all the parameterizations of 
$\rho_1(a)$ and $\rho_2(a)$ and their derivatives in this letter 
and it allows them to fit the supernova data 
(which require the densities 
to have a very small slope over the relevant portion of $a$) and 
still reach a high density in a relatively 
short time.  However, the 
relevant quantity that determines when a structure dissociates 
is $F_{inert}$, which is proportional to 
$2 \rho(a) + \rho'(a) a$.  This combination of terms is responsible 
for any local maxima in $F_{inert}$, 
not merely $\rho$ or $\rho'$.  An example of a model that has an 
inflection in $\rho_{DE}'(a)$ but none in 
$\rho_{DE}(a)$ is given by $\rho_{DE}(a) = \alpha (a - \ln[1+e^{a-C}])$, 
where $\alpha, C>0$ are constants.  
Such a model has a local maximum in $F_{inert}$ 
and if we shift by $D$, where $D > C$ is a constant, 
neither the resulting $\rho_{DE}(a+D)$ nor its derivative 
has an inflection point, but the resulting $F_{inert}$ 
still has a local maximum.

It is amusing to take examples of Model 2 which are more extreme than the ones
illustrated.  It is possible to design a pseudo-rip model such that 
disintegrations happen arbitrarily soon after the present time while still 
maintaining excellent fits to the supernova data.
The Sun may
not rise tomorrow.

This is a dramatic illustration of the fact that any amount of observational
data,
necessarily restricted to the past lightcone and necessarily with non-zero
errors, cannot predict anything mathematically about the future even one hour
hence without further assumptions. It is also a display of the difference between 
mathematics and physics: the physicist necessarily employs intuition about the
real world.

The earliest support for cosmological futures such as the big rip, little rip,
or pseudo-rip might come from the Planck satellite, which is expected to have data
around September 2012. If the dark energy equation of state emerges with
$w < -1$, it will be a shot in the arm for such exotic ideas.

It will also lend new understanding of the nature of time and
perhaps the beginning and possible cyclicity of the universe.

{\it Acknowledgements}---P.H.F. and K.J.L. were 
supported in part by the Department of Energy (DE-FG02-05ER41418).
R.J.S. was supported in part by the Department of Energy (DE-FG05-85ER40226).

\end{document}